\documentstyle[psfig]{aa}

\newcommand {\mcg} {{MCG$-$1$-$24$-$12 }}
\newcommand {\exo } {{EXO0917.3$-$0722 }}
\newcommand {\flux} {{$\times$ 10$^{-11}$ erg cm$^{-2}$ s$^{-1}$}}
\def\magcir{\raise -2.truept\hbox{\rlap{\hbox{$\sim$}}\raise5.truept
\hbox{$>$}\ }}

   \title{BeppoSAX/PDS identification of the true counterpart of the Piccinotti source H0917-074}

   \author{A.~Malizia\inst{1}, G.~Malaguti\inst{1}, L.~Bassani\inst{1}, M.~Cappi\inst{1}, 
           A.~Comastri\inst{2}, G.~Di~Cocco\inst{1}, E.~Palazzi\inst{1}, C.~Vignali\inst{3}}

   \offprints{A. Malizia (malizia@bo.iasf.cnr.it)}

   \institute{IASF/CNR, via Piero Gobetti 101, I-40129 Bologna, Italy \and 
              INAF/Osservatorio Astronomico di Bologna, via Ranzani 1, I-40127 Bologna, Italy \and
              Department of Astronomy and Astrophysics, Pennsylvania State University, 525 Davey Lab, 
              University Park, PA 16802}
   \date{Received / accepted}

   \titlerunning{BeppoSAX/PDS identification of the true counterpart of  the Piccinotti source H0917-074}

   \authorrunning{A. Malizia et al.}

\begin{document}

\abstract{High energy emission has been discovered serendipitously by
the BeppoSAX/PDS telescope in the $\sim$1.3$^{\circ}$ field of view
around the Piccinotti source H0917-074.  A re-pointing of
BeppoSAX/NFI has allowed the association of this emission
with the Seyfert 2 galaxy \mcg which lies within the original HEAO1/A2
error box of H0917-074.  This is the first PDS serendipitous discovery
of a Seyfert 2 galaxy and the first detection of \mcg in the X-ray
domain.  The measured 2-10 keV flux of  \mcg is $\sim$1 $\times$
10$^{-11}$ erg cm$^{-2}$ s$^{-1}$ compatible with the Piccinotti HEAO-1/A2
observation. This is a factor of $\sim$6 greater
than that observed from EXO0917.3-0722, originally suggested as the counterpart 
of the Piccinotti source.  The 2-10 keV spectrum of \mcg shows the presence of Fe
K$_{\alpha}$ emission together with an absorption feature at $\sim$8.7
keV.  At high energies, the Seyfert 2 still dominates and the observed
20-100 keV flux is $\sim$4 $\times$ 10$^{-11}$ erg cm$^{-2}$
s$^{-1}$.

 }

 \maketitle


\section{Introduction}
The Piccinotti sample (Piccinotti et al. 1982) is to date  the only
statistically complete sample of active galactic nuclei (AGN) in the
2-10 keV energy range. It was obtained using data from the A2 experiment
on the HEAO-1 satellite which performed a survey of 8.3 sr of the sky
(65.5\% coverage) with $|b| \ge20^{\circ}$, at a limiting flux of
3.1 $\times$ 10$^{-11}$ erg cm$^{-2}$ s$^{-1}$.  
The sample was composed of 36 AGN: 30 Seyfert galaxies, 
one starburst galaxy (M82), 4 BL Lac objects and one
QSO (3C273). \\
The BATSE instrument on board CGRO provided, for the first time, a systematic
coverage of the whole sample at higher energies (20-100 keV range,
Malizia et al. 1999) and for the brightest sources OSSE observations
in the 50-500 keV band have also been performed (McNaron-Brown et
al. 1995, Zdziarski et al. 2000).  
BeppoSAX has observed almost the whole Piccinotti sample 
in the broad band 0.1-200 keV energy range,
with the exception of the three Seyfert galaxies: IIIZW2,
MKN590 and NGC3227. \\ 
We have carried out a multiyear project to observe with BeppoSAX-NFI
the poorly studied (i.e. fainter) sources of the Piccinotti sample among which is 
the galaxy H0917-074 which was identified with the QSO \exo.  
Thanks to the BeppoSAX-PDS observation we discovered
that \exo is contaminated by another hard
X-ray source subsequently identified with the Seyfert 2 galaxy \mcg, which is the
true X-ray emitter detected by the A2 instrument. \\ 
It is worth noting that the high sensitivity of the PDS instrument (Frontera et
al. 1997) on board the BeppoSAX satellite has provided the opportunity to
increase the number of Seyfert 2s detected up to 100 keV, improving
our understanding of the high energy characteristics of this type of
object.  BeppoSAX observations of Seyfert 2 galaxies have
demonstrated that these objects can be powerful hard X-ray emitters
(above 10 keV) even though their 2-10 keV radiation is severely
attenuated by absorption in thick material (Bassani et al. 1999).
In fact, hard X-ray spectra are probably the best tool to directly measure the
absorption affecting Seyfert 2 nuclei.\\
In this paper the BeppoSAX broad band spectrum of the first PDS
serendipitous discovery of a Seyfert 2 galaxy, \mcg, is presented.
From this study, which is also the first of this source in the X-ray
domain, \mcg turns out to be a Compton thin Seyfert 2 galaxy
(N$_{H}\sim$7 $\times$ 10$^{22}$ atoms cm$^{-2}$) with a 
slightly peculiar spectrum.

\section{Observations: from \exo to \mcg}
Within the multiyear project of completing the observation of poorly
studied Piccinotti sample sources, H0917-074 was targeted by the BeppoSAX
NFI from April 15th to April 17th, 2000.  This source was 
unidentified in the original Piccinotti list and has been subsequently
associated to the QSO/Seyfert 1 galaxy EXO0917.3-0722 ($z$=0.169) by
Giommi et al. (1991) using EXOSAT observations.  More recently, the
source has been observed by ROSAT in soft X-rays (Schartel et
al. 1997) and by BATSE on-board CGRO in the hard X-ray band (Malizia
et al. 1999).  A quick-look inspection of the BeppoSAX MECS-PDS broad
band spectrum of \exo indicates an apparent inconsistency.  In fact,
as shown in figure 1, a simple power-law fit requires a MECS-PDS
normalization constant of $\sim$8-10, i.e.  an order of magnitude
greater than that normally used (Fiore, Guainazzi \& Grandi 1998).
The simplest explanation for this phenomenon is the presence of a
contaminating serendipitous source in the field of view of the PDS,
which is hexagonal in shape and 78 arcmin wide.

\begin{figure}
\psfig{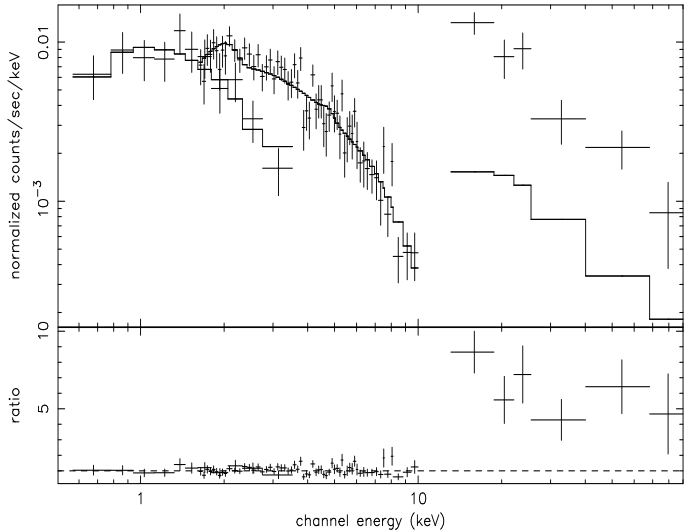}
\caption{The LECS MECS and PDS spectrum of H0917-074 (April 2000 pointing) fitted with a simple 
power law model of $\Gamma$=1.76.}
\end{figure}

The region of the sky around \exo was searched for possible high
energy sources, and the only possible candidate found was the
Seyfert 2 MCG$-1-24-12$ ($z$=0.0198), which also lies
within the original HEAO 1/A2 error box (Piccinotti et al. 1982).
In figure 2 the original A2 error box for
H0917$-$074 is shown with the position of \mcg superimposed.  
\mcg has an optical magnitude of m$_v$=15 and it is in the 2MASS
catalogue (m$_J$=14.2, m$_H$=13.3, m$_K$=12.8); 
it belongs to the 60$\mu$m sample of warm IRAS galaxies (de Grijp et
al. 1992, Kinney et al. 2000) with a log(L$_{FIR}$)=10.01L$_{\sun}$ (Chatzichristou 2000).
It is also a radio galaxy with a F$_{3.6cm}$ = 9.1 mJy and its 3.6cm radio contours show a $\sim$70 pc
elongation towards the west (Schmitt et al. 2001),
but it was never observed in the X-ray band.  
After finding \mcg as the possible candidate, we were able 
to obtain a pointed observation with BeppoSAX
and found that its 2-10 keV flux of $\sim$1 $\times$ 10$^{-11}$ erg cm$^{-2}$ s$^{-1}$ 
is more consistent with that reported in the original Piccinotti source
list (Piccinotti et al. 1982) than that observed from \exo. 
This result indicates that the Seyfert 2 is the correct identification of H0917-074.  
Also BATSE detected from this region a 20-100 keV flux of
$\sim$5 \flux (Malizia et al. 1999) which is in perfect agreement with
the PDS 20-100 keV flux of $\sim$4 \flux reported here from \mcg, suggesting that it is this 
source that was observed.

\begin{figure}
\psfig{figure=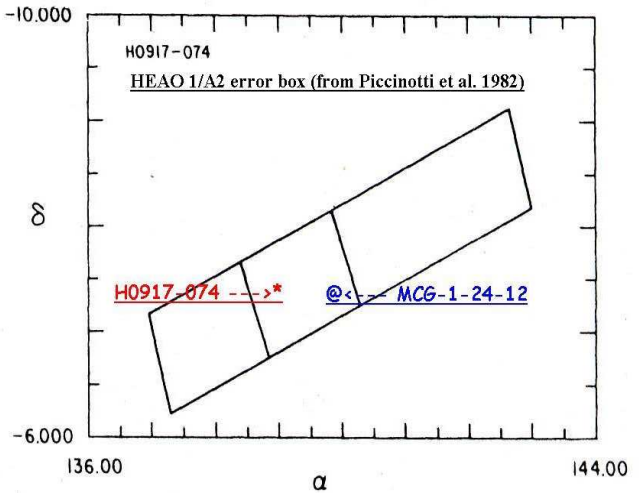,height=7.0cm,width=9cm,angle=-90}
\caption{The HEAO-1/A2 error box for H0917-074 adapted from Piccinotti et al. (1982). 
$\alpha$ and $\delta$ are in degrees (equinox 1950).
The inner and outer boxes are the 90\%
confidence boxes as described in Marshall et al. (1979). The BeppoSAX 
coordinates for H0917-074 and \mcg are superimposed.}
\end{figure}

\section{Spectral analysis}
Standard data reduction was performed using the software package
"SAXDAS" (see http://www.sdc.asi.it/software and the Cookbook for
BeppoSAX NFI spectral analysis, Fiore, Guainazzi \& Grandi 1998).
Data were linearized and cleaned for Earth occultation periods and
unwanted periods of high particle background (satellite passages
through the South Atlantic Anomaly). The LECS, MECS and PDS backgrounds
are relatively low and stable (variations of at most 30\% during the
orbit) thanks to the satellite's low inclination orbit (3.95 degrees).
Data were accumulated for Earth elevation angles $>5$ degrees and
magnetic cut-off rigidity $>6$. For the PDS data we adopted a fine
energy and temperature dependent Rise Time selection, which decreases
the PDS background by $\sim 40 \%$, improving the signal to noise
ratio of faint sources by about 1.5 (Frontera et al. 1997, 
Fiore, Guainazzi \& Grandi 1998).  Spectral fits were
performed using the XSPEC 11.0.1 software package and public response
matrices as from the 1998 November issue.  PI (Pulse Invariant)
channels are rebinned in such a way as to sample the instrument resolution with the same
number of channels at all energies when possible and to have at least
20 counts per bin.  This allows the use of the $\chi^2$ method in
determining the best fit parameters, since the distribution in each
channel can be considered Gaussian.  Constant factors have been
introduced in the fitting models in order to take into account the
inter-calibration systematic uncertainties between instruments
(Fiore, Guainazzi \& Grandi 1998).  All the
quoted errors correspond to 90\% confidence level for two interesting parameters.
The models used in what
follows contain an additional term to allow for the absorption of
X-rays due to our galaxy, which in the direction of both \exo and \mcg
corresponds to N$_{H}$ = 3.5 $\times$ 10$^{20}$ atoms cm$^{-2}$.

\subsection{\exo}
BeppoSAX pointed at \exo for an effective exposure time of
10 ks for the LECS, 52 ks for the MECS, and
25 ks for the PDS. Spectral data were extracted from
a region centred on \exo with a radius of 4 arcmin. The net count rate
was $1.67\pm0.14\times10^{-2}$ cts/s in the LECS,
$2.48\pm0.07\times10^{-2}$ cts/s in the MECS, and $0.28\pm0.03$ cts/s
in the PDS.  LECS (0.1-4 keV) and MECS (2-10 keV) data are well fitted
by a simple power law of photon index $\Gamma$ = 1.76$\pm$0.09
absorbed by the Galactic column density ($\chi^{2}$/$\nu$=90.7/91).  The
addition of an iron line is not required by the data.  If a narrow
line ($\sigma$=0) is fixed at 6.4 keV energy, the upper limit of its
equivalent width is EW$<$210 eV.  The 0.1-10 keV spectrum of \exo is
consistent with the average spectral characteristics of a sample of 10
low redshift quasars observed by BeppoSAX (Mineo et al. 2000).  The
2-10 keV flux is 1.7 $\times$ 10$^{-12}$ erg cm$^{-2}$ s$^{-1}$
(L$_X \sim$9 $\times$ 10$^{43}$ erg s$^{-1}$) which is roughly an order of magnitude less than the A2
measurement of 3.4 \flux (Piccinotti et al. 1982).

\begin{table*}
\begin{center}
\caption{\mcg: spectral analysis}
\begin{tabular}{clllllllll}
\hline
Model  & N$_{H} \times$ 10$^{22}$ & $\Gamma$              & E$_{Line}$ (keV)            & EW (eV)                & E$_{Notch}$ (keV)  & Width$_{Notch}$ (keV) & R & E$_c$ (keV)  &  $\chi^{2}$/$\nu$ \\
\hline
 (1)   & 6.27$^{+0.53}_{-0.57}$  & 1.59$^{+0.07}_{-0.03}$ & 6.40$^{+0.16}_{-0.18}$ & 176$^{+61}_{-61}$ &        -     &     -           &  -   &         & 95/62 \\
 (2)   & 6.25$^{+0.45}_{-0.45}$  & 1.57$^{+0.07}_{-0.07}$ & 6.40$^{+0.15}_{-0.20}$ & 157$^{+58}_{-61}$ & 8.70$^{+0.40}_{-0.15}$ & 0.23$^{+0.11}_{-0.11}$ &    -         &       -    & 77/60 \\
 (3)   & 6.54$^{+0.63}_{-0.61}$  & 1.74$^{+0.36}_{-0.27}$ & 6.38$^{+0.20}_{-0.18}$ & 130$^{+61}_{-55}$ & 8.72$^{+0.15}_{-0.19}$ & 0.20$^{+0.07}_{-0.12}$ & 1.62$^{+3.38}_{-1.12}$ & $>$80 & 68/58 \\
\hline
\end{tabular}
\end{center}
\end{table*}

\subsection{\mcg}
The BeppoSAX-NFIs pointed at \mcg from May 15th to May 17th, 2001.  The
effective exposure times are 28 ks for the LECS,
76 ks for the MECS, and 36 ks for the PDS.
Spectral data were extracted from a region centred on MCG$-1-24-12$
with a radius of 4 arcmin.  LECS and MECS background spectra were
extracted from blank sky fields using regions of the same size in
detector coordinates.  The net count rate was $8.5\pm0.6\times10^{-3}$
cts/s in the (1-4 keV) LECS energy band, $9.1\pm0.1\times10^{-2}$
cts/s in the (2-10 keV) MECS energy range, and $0.43\pm0.03$ cts/s in
the (20-100 keV) PDS band.  Source plus background light curves did
not show any significant variation, therefore timing analysis has
not been performed and all data have been grouped for spectral
analysis.  The main results are summarized in table 1.

\begin{figure}
\psfig{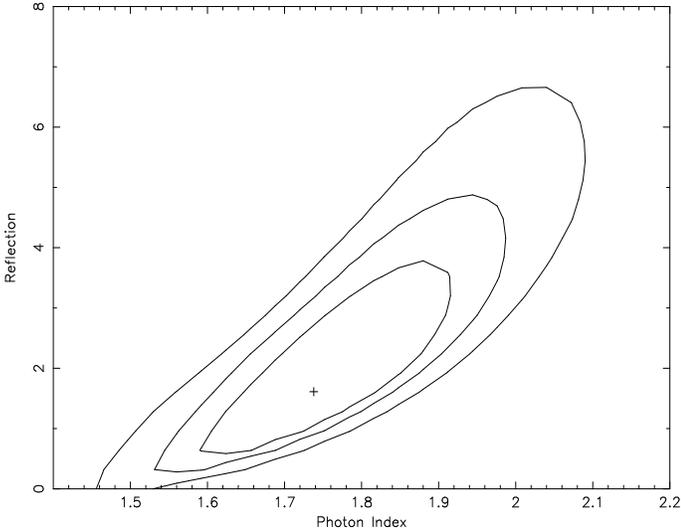}
\caption{68, 95 and 99 per cent confidence contours of reflection versus photon index
for model 3.}
\end{figure}

Due to the low statistical quality of the data at lower energies, 
only data above 1 keV have been considered.
The broad band (1-100 keV) spectrum of \mcg was firstly modeled 
with a flat, absorbed power law plus a gaussian line (model 1, table 1).
The results are $\Gamma$ = 1.59$^{+0.07}_{-0.03}$ and 
N$_{H}$ = 6.27$^{+0.53}_{-0.57}$ $\times$ 10$^{22}$ cm$^{-2}$ 
while the line is centered at 6.4$^{+0.16}_{-0.18}$ keV 
with the equivalent width of EW = 176$^{+61}_{-61}$ eV. The line is consistent
with being narrow at the 99\% confidence level, therefore we have fixed the
value of its width (sigma) to zero.  
The residuals clearly show evidence of two more features:
an absorption edge at E$\sim$8 keV and a possible
reflection hump at around 30 keV.
We have therefore considered an absorption edge 
to account for the deficit at around 8 keV . This
turns out to be located at E=8.4$^{+0.4}_{-0.6}$ keV, with optical
depth $\tau$=0.26$^{+0.17}_{-0.16}$.  The addition of this absorbed feature to
the previous model is statistically significant at $>$98\% confidence
using the F-test ($\chi^{2}$ decreases by 12 for 2 more fitting
parameters).  The presence of the feature has been confirmed 
as it is still detected when using different background event files extracted 
from blank areas in the field of view of MCG-1-24-12.  
Even taking into account the
presence of the absorption edge, the data to model ratio still shows
residuals at around 8 keV which cannot be fitted by adding one more edge.
Therefore we have replaced the edge with a notch line model 
which is equivalent to a very saturated absorption line (model 2). The quality
of the fit improves: $\Delta\chi^{2}$=6 with the same degrees of
freedom.  The energy of the absorption line is now 8.7$^{+0.4}_{-0.15}$
keV and the line width is 0.23$^{+0.11}_{-0.11}$ keV (see 3.2.1).  \\ 
For the excess at 30 keV, we take into account a possible reflection component
(model 3).
When introduced, the reflection turns out
to be strong (R=1.62$^{+3.38}_{-1.12}$) with a cut-off at E$_c>$80 keV.
The photon index is compatible with the expected canonical value
$\Gamma$=1.74$^{+0.36}_{-0.27}$ of Seyfert 2 galaxies.  In figure 3 the confidence contours
of the reflection component versus the photon index are shown.
The energies of both the emission line and the absorption
line do not change significantly from the previous ones.
In figure 4, the confidence contours of the notch
are shown for this more complex model.  
The addition of the reflection component further improves the quality of the fit
providing us with our best-fit model. In figure 5 the 1-100 keV broad band spectrum of 
\mcg fitted with model 3 is shown.

\begin{figure}
\psfig{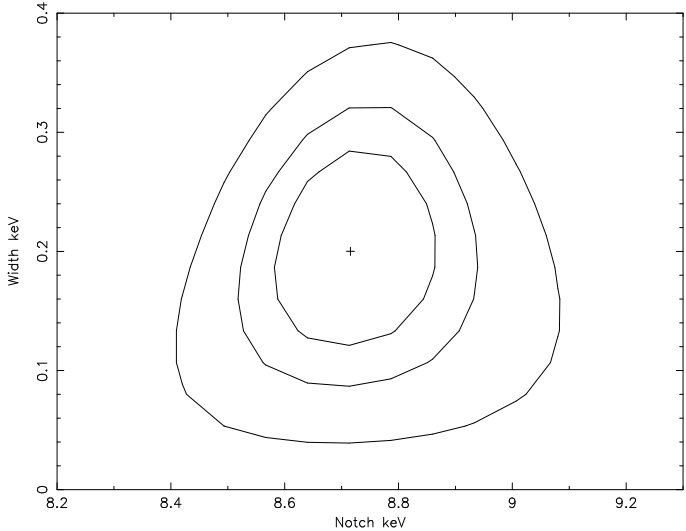}
\caption{68, 95 and 99 per cent confidence contours of the notch: width versus energy of the notch.}
\end{figure}

\begin{figure}
\psfig{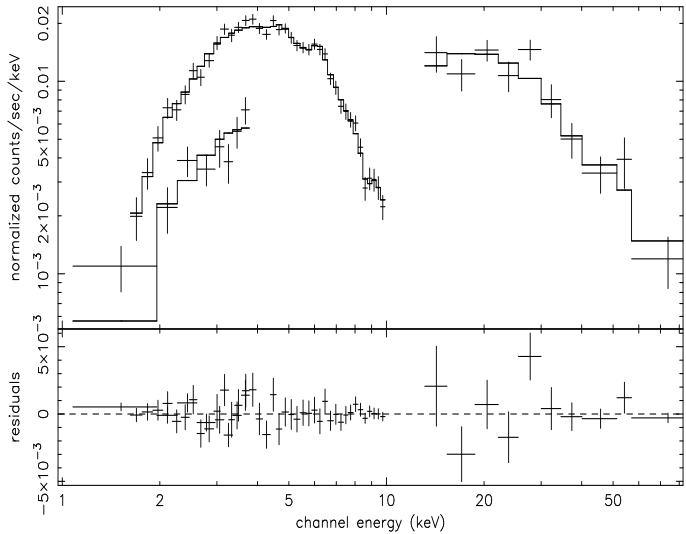}
\caption{The LECS MECS and PDS spectrum of MCG-1-24-12 fitted with model 3 
(best-fit model) in table 1.}
\end{figure}

\subsubsection{The absorption feature at $\sim$8 keV}
The best fit model that explains all the features of the 1-100 keV
spectrum of \mcg is an absorbed power law 
plus a Compton reflection component (R $\sim$ 1.6) from a cold slab
isotropically illuminated by the power law photons as in the PEXRAV
model in XSPEC by Magdziarz \& Zdziarski (1995), an Iron K$\alpha$ line at energy
E$_{line}\sim$6.4 keV and an absorption feature at energy
E$_{notch}\sim$8.7 keV.\\ 
The broad band spectrum of \mcg is
consistent with that generally found in Seyfert 2 objects,
but the detection of the absorption line makes the spectrum of
\mcg rather peculiar.  Generally, the detection of an absorption edge
indicates the presence of a warm photoionized absorber along the line of
sight; therefore we have tried to add a uniform warm absorber in this model as
described by the ABSORI model (Done et al. 1992).   
The energy of the edge ($\sim$ 8 keV)
strongly suggests that the absorption is mainly due to FeXXIII-FeXXV.
Assuming Dalgarno \& Layzer (1987) abundances, the photoelectric cross
section of neutral Fe per hydrogen atom is about 7 $\times$
10$^{-25}$; therefore, taking the optical depth to be $\sim$0.2, as observed, the
column density of the warm material (N$_{H_{W}}$ $\sim$
$\tau$/$\sigma$) is found to be $\sim$3 $\times$ 10$^{23}$ atoms
cm$^{-2}$.  \\
Fixing to this value the N$_{H_{W}}$ parameter in the ABSORI
model, the fit gives an ionization parameter $X_{i} \sim$ 290 
that is still compatible with the 
the presence of a neutral Fe line (E$_{line}\sim$6.4 keV) but 
some residuals at $\sim$8 keV still remain and 
the quality of the fit is worst ($\chi^{2}$/$\nu$ = 130/57). \\ 
Another way to explain the production of the absorption edge in the
spectrum of \mcg is by reflection from a ionized disk (PEXRIV model)
which has recently been used successfully to fit the BeppoSAX data of a
sample of Seyfert 1 objects (De Rosa et al. 2001).  As generally
observed for the Seyfert 1s, the spectrum turns out to be flat
($\Gamma \sim$ 1.64), the reflection component is R$\sim$0.4, and the
ionization parameter is around 200; however, again the quality of the fit
does not improve ($\chi^{2}$/$\nu$ = 81/55) with respect to the neutral 
reflection model.
We conclude that 
neither of these two models which allow for the presence of warm material 
at the source, provide a better fit than simple reflection
from a standard cold disk.
This may be due to the fact that
both the ABSORI and PEXRIV models are more appropriate
to fit edge-like features and not an absorption line such as that found
in our data.\\
An absorption feature at around 8 keV has been detected in a few Seyfert 
galaxies, e.g. in M81 (Pellegrini et al. 2000) and in NGC3516 
(Nandra et al. 1999, Constantini et al. 2000) which
has been explained recently in terms of the influence of resonant absorption on the iron
emission line by Ruszkowski and Fabian (2000).  However, it is worth
noting that the BeppoSAX/MECS instrument has not sufficient resolution at
these energies to resolve the line. With our data  we can conclude that
the absorption feature is produced by ions from FeXXIII
to FeXXV (8.5 - 8.9 keV) within the 90\% confidence interval, while its
best fit energy (8.7 keV) corresponds to the FeXXIV K-edge (Makishima
1985).
These values are marginally compatible with the energy of the line
at 6.38$^{+0.20}_{-0.18}$ keV.  Also the EW of the emission line and the
width of the absorption line are consistent 
within the uncertainties of the various parameters
with the measured column
density of N$_{H}$=6.5$^{+0.63}_{-0.61}$ $\times$ 10$^{22}$ cm$^{-2}$. \\
Higher quality data such as those possible with XMM-Newton
will provide definite evidence for this feature and will
help to better investigate its nature.

\section{Conclusions}
The main result of this work is that the X-ray
source corresponding to H0917-074 in the only 2-10 keV complete sample of AGN
(Piccinotti et al. 1982) is the Seyfert 2 \mcg instead of the QSO/Seyfert 1 \exo.
\mcg was never before observed in the
X-ray domain and from the present study it turns out to be a Compton
thin Seyfert 2 galaxy (N$_{H}\sim$7 $\times$ 10$^{22}$ atoms cm$^{-2}$) with an almost standard
spectrum.
The only peculiarity in this spectrum is an absorption feature at 
around 8 keV which cannot be explained by the presence of warm material around 
the source. Our best fit model is a power law absorbed by uniform cold 
material and reflected from a standard cold disk. \\
With this observation the number of Seyfert 2s belonging to the Piccinotti 
sample grows to 8 out of 30 Seyfert galaxies:
all these objects have column densities  N$_{H}>$10$^{22}$ atoms cm$^{-2}$.
In addition a few type 1 Seyferts in the Piccinotti sample have N$_{H}$
exceeding this value (NGC4151, Zdziarski et al. 2001; NGC526A, Landi et al. 2001;
NGC3783, De Rosa et al. 2002). This sample is complete down to a flux limit of
3.1 $\times$ 10$^{-11}$ erg cm$^{-2}$ s$^{-1}$ and therefore can be used to 
estimate with some confidence the percentage of absorbed sources at these 
high fluxes: out of 31 objects identified with Seyfert/QSO (Malizia et al. 1999 and this 
paper), 11 are absorbed above 10$^{22}$ atoms cm$^{-2}$, implying 
that a consistent fraction (30$\pm$12\%) of
all AGN in the 2-10 keV band are absorbed.

\begin{acknowledgements}
This research has made use of SAXDAS linearized and cleaned event files produced at
the BeppoSAX Science Data Center. We would like to thank the BeppoSAX SDC for the 
important assistance and the Mission Planning team 
for their fundamental contribution in performing the observations discussed in this work.
We would also thank J. Stephen for a careful reading of the manustript.
This research has been partially supported by ASI contracts I/R/103/00 and I/R/107/00.
CV also acknowledges the financial support by Chandra X-ray Center grant DD1-2012X
and by NASA LTSA grant NAG5-8107.
\end{acknowledgements}


\begin{thebibliography}{}
\bibitem{} Bassani, L., Cappi, M., Malaguti, G. 1999, Astr. Lett. and Communications, Vol. 39, 41
\bibitem{} Chatzichristou, E. T. 2000, ApJ, 544, 712
\bibitem{} Costantini, E., Nicastro, F., Fruscione, A., Mathur, S., Comastri, A., et al. 2000,
           ApJ, 544, 283
\bibitem{} Dalgarno, A and Layzer, D. 1987, Spettroscopy of Astrophysical Plasmas, 
           Cambridge Univ. Press., Cambridge
\bibitem{} de Grijp, M.H.K., Keel, W.C., Miley, G.K., Goudfrooij, P. and Lub J. 1992, A\&AS, 96, 389
\bibitem{} De Rosa, A., Fabian, A. C., Piro, L., Ballantyne, D. R., in the Proceedings of the 
           Symposium on `New Visions of the X-ray Universe in the XMM-Newton and Chandra Era', 
           2001, ESTEC, The Netherlands
\bibitem{} De Rosa, A., Piro, L., Fiore, F., Grandi, P., Maraschi, L., et al. 2002, A\& A, 387, 838 
\bibitem{} Done, C., Mulchaey, J. S., Mushotzky, R. F., Arnaud, K. A. 1992, ApJ, 395, 275
\bibitem{} Fiore F., Guainazzi M., Grandi P. 1998, BeppoSAX Cookbook 
\bibitem{} Frontera F., Costa E., Dal Fiume D., et al. 1997 A\&AS, 122, 357
\bibitem{} Giommi, P., Tagliaferri, G., Beuermann, K., Braduardi-Raymont, 
           Brissenden, R., et al. 1991, ApJ, 378, 77 
\bibitem{} Kinney, A. L., Schmitt, H. R.,  Clarke, C. J., Pringle, J. E., Ulvestad, J. S., 
           Antonucci, R. R. J. 2000, ApJ, 537, 152
\bibitem{} Landi, R.; Bassani, L.; Malaguti, G.; Cappi, M.; Comastri, A.;
           Dadina, M.; et al. 2001, A\%A 379, 46
\bibitem{} Magdziarz, P.; Zdziarski, A. 1995, MNRAS, 273, 837
\bibitem{} Makishima, K. 1985, In: Mason K.O., Watson M.G., White N.E. (eds.)
           The Physics of Accretion onto Compact Objects, Springer-Verlag, Berlin, p. 249
\bibitem{} Malizia, A., Bassani, L., Zhang, S. N., Dean, A. J., Paciesas, W. S., Palumbo, G. G. C.
           1999, ApJ, 519, 637
\bibitem{} Marshall, F. E., Boldt, E. A., Holt, S. S., Mushotzky, R. F.,,Rothschild, R. E., et al. 1979,
           ApJS, 40, 657
\bibitem{} McNaron-Brown, K., Johnson, W. N., Jung, G. V., Kinzer, R. L., Kurfess, J. D., Strickman, M. S.,
           et. al. 1995, ApJ, 451, 575
\bibitem{} Mineo, T., Fiore, F., Laor, A., Costantini, E., Brandt, W. N., et al. 2000, A\&A, 359, 471
\bibitem{} Nandra, K., George, I. M., Mushotzky, R. F., Turner, T. J. and Yaqoob, T. 1999, ApJ, 523L, 17 
\bibitem{} Pellegrini, S., Cappi, M., Bassani, L., Malaguti, G., Palumbo G.G.C. and Persic, M.
           2000, A\&A, 353, 447 
\bibitem{} Piccinotti, G., Mushotzky, R. F., Boldt, E. A., Holt, S. S., Marshall, 
           F.~E., Serlemitsos, P.~J., and Shafer, R.~A. 1982, ApJ, 253, 485
\bibitem{} Ruszkowski, M. and Fabian, A. C. 2000, MNRAS, 315, 223
\bibitem{} Schartel, N., Schmidt, M., Fink, H. H., Hasinger, G., and Trumper, J. 1997, A\&A, 495, 749
\bibitem{} Schmitt, H. R. and Kinney, A. L. 2001, ApJS, 128, 479
\bibitem{} Zdziarski, A. A., Poutanen, J., Johnson, W. N. 2000, ApJ, 542, 703 
\bibitem{} Zdziarski, A. A., Leighly, K. M., Matsuoka, M., Cappi, M., Mihara, T. 2001, {\it astro-ph} 0108036
\end{thebibliography}
\end{document}